\documentclass[prd,twocolumn,twoside,preprintnumbers,superscriptaddress,nofootinbib,natbib,bm,amsrefs]{revtex4-2}
\usepackage{graphicx}
\usepackage{url}
\usepackage{xcolor}
\usepackage{amsmath, amstext, amsfonts,amssymb, amsthm}
\usepackage{comment}
\usepackage[utf8]{inputenc}
\usepackage{multirow}

\newcommand{\ov}{\overline}

\newcommand{\eg}{{\em e.g.}\,\,}

\definecolor{BlueViolet}{rgb}{0.2, 0.00, 0.7}
\definecolor{Blue}{rgb}{0.15, 0.00, 0.9}
\definecolor{lightblue}{rgb}{0.15, 0.35, 0.95}
\definecolor{kitgreen}{rgb}{0,
0.58823 
, 0.50980 
}
\usepackage[
colorlinks=true, linkcolor=lightblue,citecolor=lightblue,urlcolor=kitgreen]{hyperref}

\newcommand{\Eprint}[1]{\href{#1}}

\definecolor{lb}{rgb}{.74,.83,.9}
\definecolor{ly}{rgb}{1,.92,.8}
\definecolor{lr}{rgb}{.98,.85,.87}

\begin{document}

\title{\includegraphics[width=0.55\textwidth]{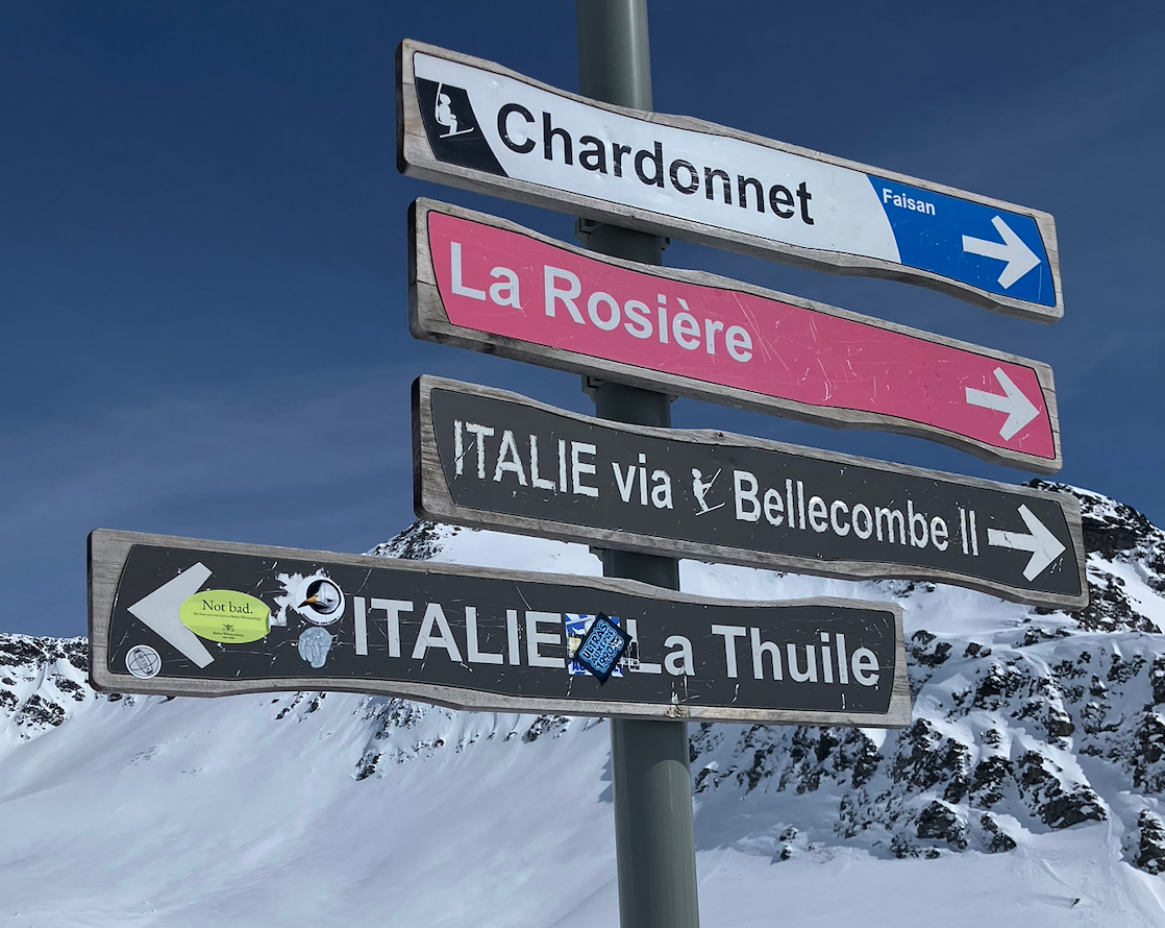}
\\[1em]
$b \to c$ semileptonic sum rule: SU(3)$_{\rm{F}}$ symmetry violation}

\author{Syuhei Iguro}
\affiliation{Institute for Advanced Research, Nagoya University, Nagoya 464-8601, Japan}
\affiliation{Kobayashi-Maskawa Institute for the Origin of Particles and the Universe, Nagoya University, Nagoya 464--8602, Japan}

\begin{abstract}
To clarify possible deviations in $b\to c\tau\overline\nu$ processes, the $b\to c$ semileptonic sum rule provides a valuable tool.
This relation, derived based on heavy quark symmetry (HQS), offers a powerful consistency check among experimental results.
In this work, we extend the previously proposed sum rule for $\{B\to D^{(*)} l\overline\nu,\,\Lambda_b\to \Lambda_c l\overline\nu\}$ to include $\{B_s\to D_s^{(*)} l\overline\nu,\,\Xi_b\to \Xi_c l\overline\nu\}$, thereby enabling more useful cross-checks.
Although the relation is supported by HQS and SU(3) flavor symmetry, both symmetries are broken in reality, and the size of the violation needs to be quantified to assess the validity of the sum rule.
While the violation is expected to be moderate based on chiral perturbation theory, we perform a numerical evaluation and compare it with future experimental sensitivities.
We find that the violation remains smaller than the expected experimental uncertainty.
Therefore another new physics agnostic and predictive sum rules are constructed to check the consistency.\\
---------------------------------------------------------------------------------------------------------------------------------
\end{abstract}
\maketitle

\section{Introduction}
\label{sec:intro}

The heavy quark symmetry (HQS) \cite{Isgur:1989vq,Isgur:1990yhj,Neubert:1993mb} and SU(3) flavor symmetry (SU(3)$_{\rm{F}}$) \cite{Gell-Mann:1961omu,Neeman:1961jhl,Gell-Mann:1962yej} are approximate symmetries of the Standard Model (SM). 
They offer tools for theoretical calculations in heavy hadron decays to determine fundamental SM parameters and probe the new physics (NP).
They are exact in the limit of an infinite heavy quark mass and the degenerate masses of light quarks where heavy quarks and light quarks respectively correspond to $Q=b,\,c$ and $q=u,\,d,\,s$.
The former symmetry especially plays an important role in singly heavy flavored hadron transitions \eg a beauty hadron ($H_b$) decays into a charming hadron ($H_c$).
The typical size of the HQS violation in the $H_b\to H_c$ decay system is estimated to be of order $\epsilon_Q=\Lambda_{\rm QCD}/(2m_Q)\simeq 10\sim20\%$ where $\Lambda_{\rm QCD}$ is a parameter of the QCD scale.
The heavy quark effective theory (HQET) allows us to expand the hadronic transition form factors (FFs) in $\epsilon_Q$.
This framework successfully describes most of the existing data well \cite{PDG2024}.
On the other hand, chiral perturbation theory suggests that SU(3)$_{\rm{F}}$ symmetry works well in $H_b\to H_c$ transitions \cite{Laiho:2005ue}.
This has been confirmed for $B\to D^{(*)}_{(s)}$ transition FFs at about $5\sim10\%$ \cite{McLean:2019qcx,Harrison:2023dzh}.
Such confirmation is not solid for baryon decays since there is no Lattice calculation for $\Xi_b\to\Xi_c$ transition.
Another source of violation arises from the hadron masses.
The mass differences among bottomed hadrons and charmed hadrons respectively are about $5\%$ and $10\%$ which also affects kinematics \eg the phase space range.
On the other hand, experimental data show no significant deviation from the expectation such as
$\Gamma(B^0\to Dl\ov\nu)/\Gamma(B_s\to D_sl\ov\nu)=1.10\pm0.10$ and $\Gamma(B^0\to D^*l\ov\nu)/\Gamma(B_s\to D_s^*l\ov\nu)=1.07\pm0.10$ \cite{PDG2024}

In recent years a $4\sigma$ level deviation in $b\to c\tau\ov\nu$ triggered theorists to propose an HQET based sum rule among a measure of lepton flavor universality violation, $R_{H_c}={\rm{BR}}(H_b\to H_c\tau\ov\nu)/{\rm{BR}}(H_b\to H_c l\ov\nu)$ where $l$ being $e,\,\mu$,\footnote{In the conference, BaBar announced a new $R_{D^{(*)}}$ measurement based on the semileptonic tagging method. 
$R_{D^*}$ is measured to be slightly smaller and consistent with the SM prediction within $2\sigma$ for $R_{D^{(*)}}$.
The result is away from their previous result without a clear explanation during the conference, making the situation unclear.} 
\begin{align}
      \frac{R_{\Lambda_b}}{R_{\Lambda_b}^{\rm{SM}}}- \frac{1}{4} \frac{R_{D}}{R_{D}^{\rm{SM}}}-\frac{3}{4} \frac{R_{D^*}}{R_{D^*}^{\rm{SM}}}\simeq \delta\,,
      \label{eq:RSMHQw1}
\end{align}
which works as an independent cross check \cite{Endo:2025fke,Endo:2025lvy}.\footnote{See Refs.~\cite{Blanke:2018yud,Blanke:2019qrx,Fedele:2022iib,Duan:2024ayo} for the earlier studies and Refs.~\cite{Endo:2025cvu,Endo:2025set} for the relevant extensions.
Also the HQS motivates us to replace $R_{\Lambda_c}$ be $R_{X_c}$ where $X_c$ means the inclusive channel. }
A non-zero $\delta$ which stems from the potential NP, is confirmed to be negligible compared to the current experimental uncertainty \cite{HFLAV:2024ctg,LHCb:2022piu}.
One can substitute the experimental value in the numerator to predict others and check the experimental consistency.

In the presence of approximate SU(3)$_{\rm{F}}$ symmetry, it can be interesting to construct a set of sum rules involving the SU(3)$_{\rm{F}}$ rotated decays, $\{B\to D^{(*)} l\ov\nu \leftrightarrow B_s\to D_s^{(*)} l\ov\nu \}$ and $\{\Lambda_b\to \Lambda_c l\ov\nu \leftrightarrow \Xi_b\to \Xi_c l\ov\nu\}$.
Naively we can expect that the SU(3)-extended sum rules hold approximately as in Eq.~(\ref{eq:RSMHQw1}), thanks to the approximate SU(3)$_{\rm{F}}$ symmetry of the form factor and relatively small violation in the mass differences.
However, since the sum rules involve three $R$ observables and rely on cancellations among them, it is not trivial to guess how small the violation $\delta$ can be.
Furthermore the relevant $R$ observables are nice targets for the future measurements.
Table~\ref{tab:prospect} summarizes the current and projected experimental uncertainties of the $R$ observables.
The current (future) relative uncertainty is taken from Ref.~\cite{LHCb:2022piu} (\cite{Bernlochner:2021vlv}) for $\Lambda_c$, Ref.~\cite{HFLAV:2024ctg} (\cite{ATLAS:2025lrr}) for $D$ and $D^*$, and  Ref.~\cite{Bernlochner:2021vlv} for $D_s^{(*)}$ future prospect.
To our best knowledge there is no prospect available for $R_{\Xi_c}$.
For $D_s^{(*)}$, due to the limited statistics, the estimation is made by combining $D_s$ and $D_s^*$ modes.
For the future prospect, the number in the parentheses corresponds to the optimistic systematic uncertainty case while the other is pessimistic case for $R_{\Lambda_c}$ and $R_{D_s^{(*)}}$.
It is seen that there will be precise data available in future.  
The SU(3)$_{\rm{F}}$-extended sum rules would provide a further motivation for future measurements at LHCb \cite{LHCb:2018roe} and FCC-ee \cite{FCC:2025lpp}.

\begin{table}[t]
\begin{center}
  \begin{tabular}{cccccc} 
Mode & $\Lambda_c$& $D$ & $D^*$ & $\Xi_c$ & $D_s^{(*)}$  \\ \hline
Current &$31\%$&$6.7\%$ &$3.9\%$&$-$&$-$\\ 
Prospect &$5\,(2)\%$&$1.3\%$&$1.0\%$&$-$&$5\,(2)\%$\\\hline 
  \end{tabular}
  \caption{
  Summary table of current and mid-term future experimental relative uncertainty in the $R_{H_c}$ observable.  
 }
 \vspace{-.4cm}
  \label{tab:prospect}
\end{center}   
\end{table}

It is natural to ask whether there is an advantage in combing various SU(3) rotated modes over simply comparing the SU(3)$_{\rm F}$ rotated modes \eg $R_D/R_D^{\rm SM}$ vs. $R_{D_s}/R_{D_s}^{\rm SM}$.
An explicit construction of the sum rules between $\Lambda_c$ and $D_s^{(*)}$ modes as well as $\Xi_c$ and $D_s^{(*)}$ modes allows the simpler comparison among relevant decays and can convey the direct message if a good sum rule is obtained.
Numerically checking this is the goal of the work.

The outline of this work is given as follows:
In Sec.~\ref{sec:FW}, we introduce our framework and then we investigate corrections to the sum rule and discuss phenomenological implications in Sec.~\ref{sec:pheno}.
We draw our conclusions in Sec.~ \ref{sec:conc}.

\section{Frame work}
\label{sec:FW}

Assuming that NP contributes only to the $b\to c \tau\bar\nu$ transitions, the weak effective Hamiltonian is given as, 
\begin{align}
 \label{eq:Hamiltonian}
 {\cal {H}}_{\rm{eff}}= 2 \sqrt2 \, G_FV_{cb}\biggl[ (1+&C_{V_L})O_{V_L}+C_{S_L}O_{S_L}\\\nonumber 
 +&C_{S_R}O_{S_R}+C_{T}O_{T}\biggl]\,.
\end{align}
We consider dimension-six effective operators given by,
\small{\begin{align}
 &O_{V_L} = (\overline{c} \gamma^\mu P_Lb)(\overline{\tau} \gamma_\mu P_L \nu_{\tau})\,,\,\,\, 
 O_{S_R} = (\overline{c} P_Rb)(\overline{\tau} P_L \nu_{\tau})\,, \label{eq:operator}\nonumber \\
 &O_{S_L} = (\overline{c} P_L b)(\overline{\tau} P_L \nu_{\tau})\,,\,\,\, O_{T} = (\overline{c}  \sigma^{\mu\nu}P_Lb)(\overline{\tau} \sigma_{\mu\nu} P_L \nu_{\tau}) \,,
\end{align}}
where $P_{L(R)}=(1\mp\gamma_5)/2$ is a chirality projection operator. 
The NP contribution is captured in the Wilson coefficients (WCs) of $C_X$ with $X$ being $V_{L}$, $S_{L,R}$, and $T$.
They are normalized to the SM contribution with a factor of $2 \sqrt2 G_FV_{cb}$ and the SM limit corresponds to $C_{X} = 0$. 
We also assume that the neutrinos are left-handed.\footnote{The violation of the sum rule Eq.~(\ref{eq:RSMHQw1}) in the presence of a massive right-handed neutrino is confirmed to be small \cite{Iguro-Kretz:2026}.
See Refs.~\cite{Iguro:2018qzf,Robinson:2018gza,Babu:2018vrl, Mandal:2020htr,Penalva:2021wye,Datta:2022czw} for the BSM models in this direction.}

\begin{figure*}[t]
\begin{center}
\includegraphics[width=0.3\textwidth]{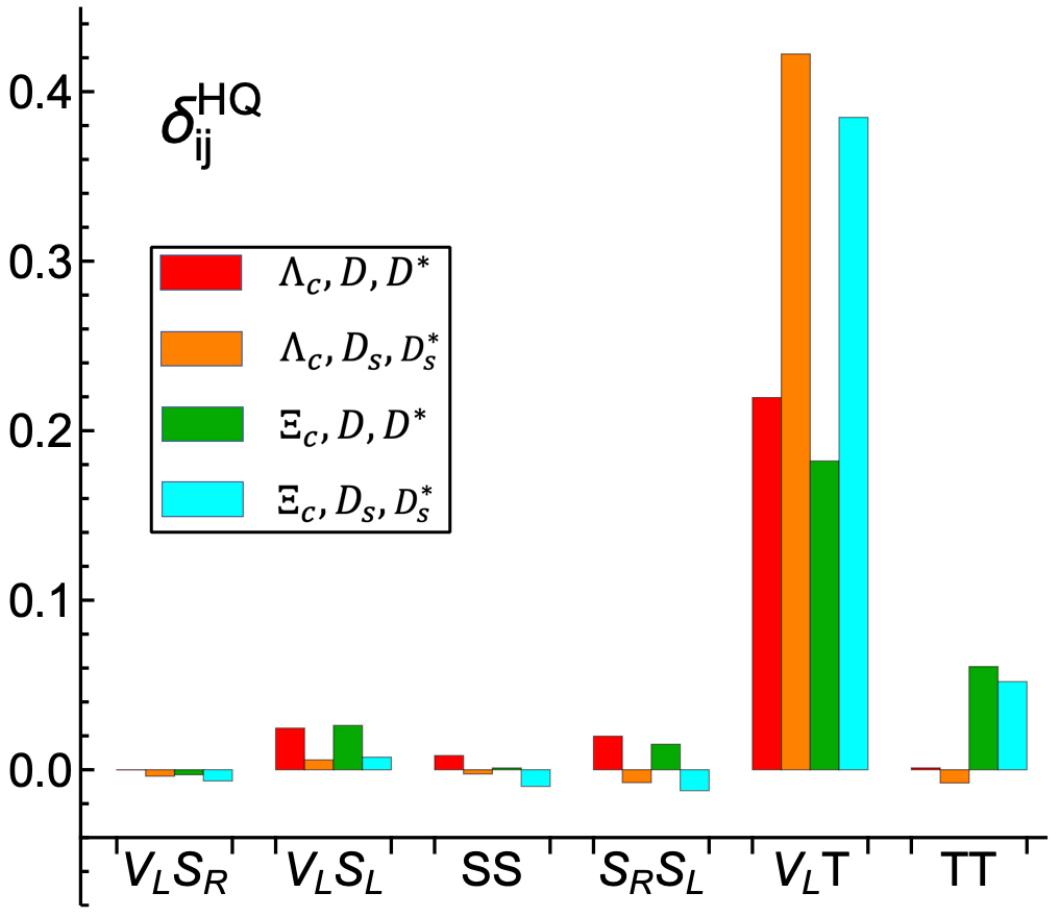}~~~~~~~~
\includegraphics[width=0.3\textwidth]{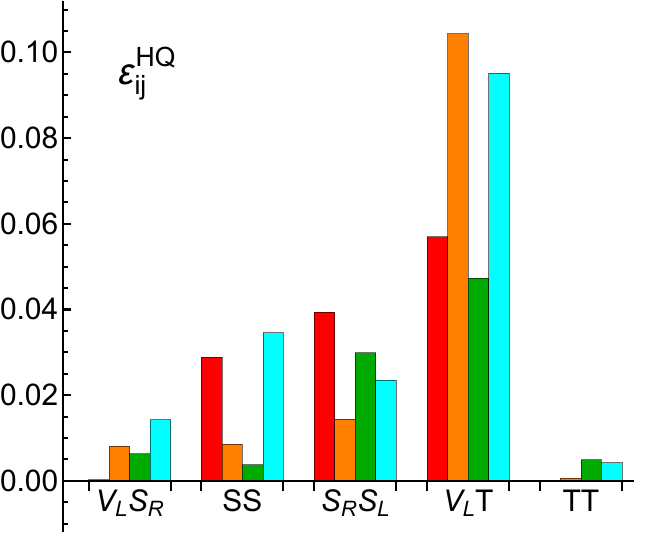}
\caption{
   \label{Fig:HQ_delep}
$\delta_{ij}$ (left) and $\epsilon_{ij}$ (right) in the HQ method.
From left to right we have $ij=V_LS_R,\,V_LS_L,\,SS,\,S_RS_L,\,V_LT,\,TT$.
Combinations of baryon and meson are shown in plot legend.
} 
\end{center}
\vspace{-3mm}
\end{figure*}

Let us study the decay rates in the heavy quark limit, $m_{c,b} \gg \bar \Lambda$ where $\bar \Lambda$ is a parameter of the QCD energy scale.
Since the heavy quark symmetry is restored in the limit, the hadron transition FFs are described  by the leading order Isgur-Wise (IW) functions in the HQET, and their corrections are suppressed.
Besides, when we deal with a static color source of the heavy quark which couples to the light quarks, the heavy quark expansion leads to \cite{Falk:1992wt,Falk:1992ws},
\begin{align}
 m_{H_Q} = m_Q\left( 1+ \bar \Lambda/m_Q + \dots \right) \,,
 \label{eq:mass_spectrum}
\end{align}
where $\bar\Lambda$ governs the light degree freedom and parameter of $\mathcal{O}(\Lambda_{\rm QCD})$. 
Therefore in the heavy quark limit, the hadron masses converge as,
\begin{align}   
 m_b=&\,m_B=m_{\Lambda_b}=m_{B_s}=m_{\Xi_b}\,,\nonumber\\     
 m_c=&\,m_{D}=m_{D^*}=m_{\Lambda_c}=m_{D_s}=m_{D_s^*}=m_{\Xi_c} \,.
 \label{eq:mass_HQL}
\end{align}
The $b\to c$ hadronic matrix elements are described by a set of FFs.\footnote{See for instance Ref.~\cite{Endo:2025fke} for an explicit formula.}
In the heavy quark limit, FFs are described by the leading order IW function, $\xi_{(s)}$ and $\zeta_{(s)}$ for $B\to D^{(*)}\,(B_{s}\to D^{(*)}_{s})$ and $\Lambda_b\to\Lambda_c\,(\Xi_b\to\Xi_c)$ transitions, respectively.
The higher order terms in heavy quark and $\alpha_s$ expansions can be systematically incorporated.
Specifically we adopt the result of Refs.~\cite{Iguro:2020cpg}, \cite{Bordone:2019guc}, \cite{Bernlochner:2018kxh}, \cite{Ebert:2006rp} respectively for $B\to D^{(*)}$, $B_s\to D_s^{(*)}$, $\Lambda_b\to\Lambda_c$ and $\Xi_b\to\Xi_c$ where the FFs are fitted at $\mathcal{O}(1/m_b,\,1/m_c^2,\,\alpha_s)$, $\mathcal{O}(1/m_b,\,1/m_c^2,\,\alpha_s)$, $\mathcal{O}(1/m_b,\,1/m_c^2,\,\alpha_s/m_Q)$ and $\mathcal{O}(1/m_Q)$\footnote{We will discuss more the $\Xi_b\to\Xi_c$ form factor in appendix \ref{app:FF}.} for each.

In the heavy quark limit, thanks to these simplifications we have the exact relation among differential decay rate as \cite{Endo:2025set},
\begin{align}
\label{eq:kappa_SR1}
    \frac{\kappa_{\Lambda_c}}{\zeta(w)^2}&=\frac{2}{w+1}\frac{\kappa_{D}+\kappa_{D^*}}{\xi(w)^2}=\frac{\kappa_{\Xi_c}}{\zeta_s(w)^2}=\frac{2}{w+1}\frac{\kappa_{D_s}+\kappa_{D_s^*}}{\xi_s(w)^2},
\end{align}
where $\kappa_{H_c}=d\Gamma(H_b \to H_c\tau\ov\nu )/dw$ is defined.
The kinematic variable of the recoil energy is introduced as $w=(m_{H_b}^2+m_{H_c}^2-q^2)/(2m_{H_b}m_{H_c})$ with the squared invariant mass of the lepton system, $q^2=(p_l+p_\nu)^2$.
We note that $\kappa_{H_c}$ is not limited to the SM operators.
By dividing the generic differential decay rate by the corresponding SM one, we obtain as,
\begin{align}
\label{eq:kappa_ratio}
 \frac{\kappa_{\Lambda_c}}{\kappa_{\Lambda_c}^{\rm SM}}=\frac{\kappa_D+\kappa_{D^*}}{(\kappa_D+\kappa_{D^*})^{\rm SM}}
 =\frac{\kappa_{\Xi_c}}{\kappa_{\Xi_c}^{\rm SM}}=\frac{\kappa_{D_s}+\kappa_{D_s^*}}{(\kappa_{D_s}+\kappa_{D_s^*})^{\rm SM}}
\,. 
\end{align}
Given the measurement is carried out in the form of $R$ observables to have a better control on the systematic uncertainty, we replace $\kappa_{H_c}$ by $\Gamma_{H_c}\equiv\int_1^{w_{\rm max}} \kappa_{H_c} dw$ and consider the following quantity \cite{Endo:2025fke},
\begin{align}
 \delta[B,\,M] \equiv \frac{R_B}{R_B^{\rm{SM}}}- \alpha \frac{R_{M_1}}{R_{M_1}^{\rm{SM}}}-\beta \frac{R_{M_2}}{R_{M_2}^{\rm{SM}}}\,,
\end{align}
where $\alpha+\beta=1$ is satisfied, and $B$ and $M_i$ correspond to labels of the charmed baryon and meson states.
Here $M=\left(M_2,\,M_1\right)^T$ is a heavy quark doublet.
When $\delta$ is small, the relation becomes more predictive.
Assuming the SM interactions ($C_X=0$), we see $\delta=0$ thanks to $\alpha+\beta=1$. 
For this construction we assumed that the NP enter only in semitauonic decays as defined in Eq.~(\ref{eq:Hamiltonian}) and normalized both numerator and denominator by the decay width of the light lepton mode.    
The generic formulae of $R_X/R^{\rm SM}_X=\sum_{ij} a_X^{ij} C_iC_j^*$ are given in appendix \ref{app:GF}.\footnote{Since $a_X^{S_RS_R}=a_X^{S_LS_L}$ holds we express them as $a_X^{SS}$ for simplicity.}
It is observed that the difference of coefficients is about $20\%$ at most implying that SU(3)$_{\rm{F}}$ works well as a guiding principle.
By taking the ratio as $R_{X_c}/R_{X_c}^{\rm SM}=\Gamma(H_b\to H_c \tau\ov\nu)/\Gamma(H_b\to H_c \tau\ov\nu)^{\rm SM}$, a fraction of the corrections is canceled and the difference becomes mild.

There are two proposals which work well for ground to ground state transitions \cite{Iguro:2026}, about how to fix the sum rule coefficient $\alpha$:
\begin{itemize}
    \item{
    Heavy quark (HQ) method \cite{Endo:2025lvy}: $\alpha=1/4$ motivated by heavy quark limit and zero-recoil limit where $\gamma_D:\gamma_{D^*}=1:3$ holds. 
    } 
    \item{
    Intuitive method developed at KIT \cite{Blanke:2018yud,Blanke:2019qrx,Fedele:2022iib}: Tuning such that the $kl$ operator combination vanishes from $\delta$ with $\alpha_{kl}=\left(a_{B}^{kl}-a_{M_2}^{kl} \right)/\left(a_{M_1}^{kl}-a_{M_2}^{kl} \right)$.
    We call this method as KIT method and show the result of this method in appendix \ref{app:KIT}.
    }
\end{itemize}

We can decompose the sum rule violation as $\delta_{\rm NP}=\sum_{ij} C_iC_j^* \delta_{ij}$.
Besides, to assess a goodness of the sum rule, we consider the cancellation measure $\epsilon$ defined as \cite{Iguro:2026},
\begin{align}
 \epsilon_{ij}= \frac{|\delta_{ij}|}{ {\rm{Max}}\bigg[\,\,\bigg|\frac{R_{X_1}}{R_{X_1}^{\rm SM}}\bigg|_{ij},\,\bigg|\alpha \frac{R_{X_2}}{R_{X_2}^{\rm SM}}\bigg|_{ij},\,\bigg|\beta \frac{R_{X_3}}{R_{X_3}^{\rm SM}}\bigg|_{ij} \,\,\bigg] }\,\,.
\end{align}
The smaller $\epsilon_{ij}$, the more precise cancellation occurs.

\section{violation and implications}
\label{sec:pheno}

\begin{table}[b]
\begin{center}
  \begin{tabular}{cccc} 
 Scenario & Parameter & Value   \\ \hline
$S_L$ & $C_{S_L}$ & $-0.57\pm0.86\,i$ \\
$S_R$ & $C_{S_R}$ & 0.18 \\
$T$ & $C_{T}$ & $0.02 \pm 0.13\,i$ \\
${\rm{R}}_2$ & $C_{S_L}=8.4\,C_T$ & $-0.09 \pm 0.56\,i$ \\
${\rm{S}}_1$ & $C_{S_L}=-8.9\,C_T$ & $0.18$ \\
${\rm{U}}_1$ & $C_{V_L}$,\,$\phi$ & $0.075,\,\pm 0.466\pi$ \\ \hline
\end{tabular}
  \caption{Fit results for WCs in single-operator ($S_L$, $S_R$, $T$) and single leptoquark (${\rm R}_2$, ${\rm S}_1$, ${\rm U}_1$) scenarios.
  Each column indicates the scenario, parameter and fitted WCs at $\mu_b$.
 }
  \label{Tab:NPscenario}
\end{center}   
\vspace{-.15cm}
\end{table}

Having established the framework, we now evaluate the size of the sum rule violation $\delta_{ij}$ and cancellation measure $\epsilon_{ij}$.
Fig.~\ref{Fig:HQ_delep} shows $\delta_{ij}$ (left) and $\epsilon_{ij}$ (right) within the HQ method.
Red, orange, green and cyan bars correspond to the $\Lambda_c$-$D$-$D^*$, $\Lambda_c$-$D_s$-$D_s^*$, $\Xi_c$-$D$-$D^*$ and $\Xi_c$-$D_s$-$D_s^*$ combinations where baryon and meson combination is given as $B$-$M_1$-$M_2$.
From left to right we consider $ij=V_LS_R,\,V_LS_L,\,SS,\,S_RS_L,\,V_LT,$ and $TT$.
It is observed that the sum rule violation is less than $0.1$ except for the $V_LT$ term for all combinations.
$\Lambda_c$-$D_s$-$D_s^*$ and $\Xi_c$-$D_s$-$D_s^*$ combinations have larger violation of $\delta^{\rm HQ}_{V_LT}\simeq 0.4$.
Different from ground to orbitally excited combinations such as $\Lambda_c$-$D_1$-$D^*_2$ and $\Lambda_c$-$\Lambda_c^*(1/2^-)$-$\Lambda_c^*(3/2^-)$ \cite{Iguro:2026}, the size of the violation is moderate and does not exceed unity. 
The maximal size of the cancellation measure is about $0.1$ which again is much smaller than that of ground to excited combinations.
We observe a better cancellation in the $TT$ term for all four combinations in the most right bins on the $\epsilon_{ij}^{\rm HQ}$ plot.
As is expected from the left plot the $V_LT$ term has the larger cancellation measure.
It is difficult to further identify a pattern of which combination has a lager violation and better cancellation.

In reality, the extent of the sum rule violation also depends on the values of WCs of NP scenarios.
Table~\ref{Tab:NPscenario} summarizes the fitted WCs for simplified NP scenarios motivated by the $R_{D^{(*)}}$ anomaly \cite{Iguro:2024hyk}.
There are three ``single operator'' scenarios and three ``single leptoquark (LQ)'' scenarios.\footnote{The ${\rm U}_1$ leptoquark scenario corresponds to the $U(2)$-flavored scenario with the relation of $C_{S_R} = -3.7e^{i\phi}C_{V_L}$.
See Ref.~\cite{Iguro:2024hyk} for further details of the fit.
}

\begin{figure}[t]
\begin{center}
\includegraphics[width=0.3\textwidth]{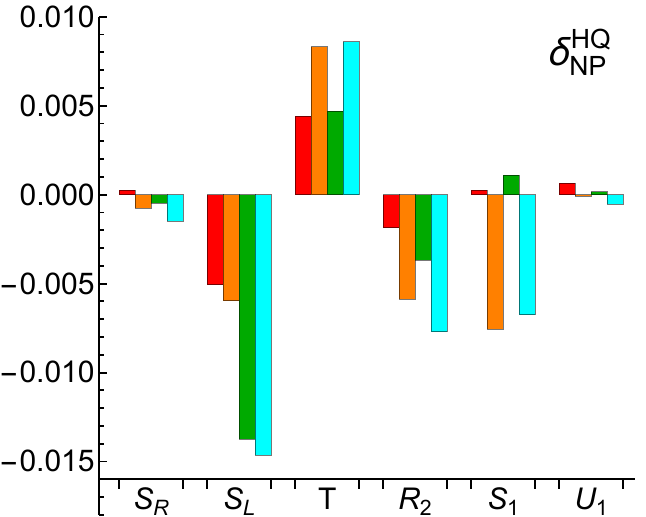}
\caption{
   \label{Fig:HQ_delNP}
$\delta_{\rm{\rm NP}}$ for six benchmark NP solutions to the $b\to c\tau\ov\nu$ anomaly in the HQ method.
Three single operator scenarios and leptoquark scenarios are shown. 
See Fig.~\ref{Fig:HQ_delep} for the color to combination correspondence.
\vspace{-4mm}
} 
\end{center}
\end{figure}

Figure~\ref{Fig:HQ_delNP} shows $\delta_{\rm NP}$ in six benchmark NP scenarios.
The color scheme is the same as Fig.~\ref{Fig:HQ_delep}.
The size of the violation is observed to be less than $\mathcal{O}(1)\%$ for the $\Lambda_c$-$D_{(s)}$-$D_{(s)}^*$ combination while this could be slightly enhanced with the $\Xi_c$-$D_{(s)}$-$D_{(s)}^*$ combination especially in the $S_L$ scenario.
However this is smaller than the expected relative experimental uncertainty of $R_{D_s^{(*)}}$ and $R_{\Lambda_c}$.
Based on the measured $R_{D_{(s)}^{(*)}}$ we can predict $R_{\Xi_c}$ in a model agnostic way.
This would motivate precise measurement of $R_{\Xi_c}$.
Besides, a consistency check between  $R_{D_{s}^{(*)}}$ and $R_{\Lambda_c}$ would be interesting.

Although we do not show the plot in the main text, let us summarize the observations within the KIT method. 
For the KIT method we can select and eliminate one operator from the sum rule violation $\delta$.
We tried all combinations and found that the $ij=V_LS_R$ case has the smallest $\delta_{\rm NP}$.
It is observed that the result of the KIT method is very similar to that of the HQ method thanks to the sum rule coefficient approximately satisfies $\alpha\sim0.25$.
We see that both methods work quite well in the ground to ground modes even with SU(3)$_{\rm{F}}$ rotations and hence the resulting sum rules are very predictive.

\section{Conclusion}
\label{sec:conc}
Based on approximate heavy quark symmetry and SU(3) flavor symmetry, we constructed sum rules connecting the decay rates of $B\to D^{(*)} l\overline\nu$, $\Lambda_b\to \Lambda_c l\overline\nu$, $B_s\to D_s^{(*)} l\overline\nu$, and $\Xi_b\to \Xi_c l\overline\nu$.
While these relations are exact in the heavy-quark limit, realistic comparisons with experimental data require the inclusion of higher-order corrections.
We have numerically evaluated the coefficients in $R_{X_c}/R_{X_c}^{\rm SM}$ formulae where $X_c=D_{(s)}^{(*)},\,\Lambda_c$ and $\Xi_c$ and found that they remain within $\sim20\%$.
Furthermore the resulting violations are significantly suppressed in the sum rules.
As a result, the predicted deviations due to NP scenarios remain below the expected experimental sensitivity.
We considered both HQ and KIT construction methods and they yield consistent results, indicating that the sum rules are robust and provide reliable experimental cross-checks.

Currently large potential uncertainty lays in the $\Xi_b\to\Xi_c$ form factor where neither $w$ distribution measurement nor Lattice calculation available.
In this work we relied on the model calculations while such a work is necessary to validate the cross check further, 
along with the estimation of experimental prospect.
When some of them become available evaluating the uncertainty of the sum rule as is done for the $\Lambda_c$-$D$-$D^*$ combination in Ref.~\cite{Endo:2025fke} should be pursued.

\section*{Acknowledgements}
I would like to thank the organizers for the generous support and hospitality.
I thank Hiroyasu Yonaha and Abhijit Mathad for the fruitful discussions and encouraging this project.
I also appreciate Motoi Endo, Satoshi Mishima, Ryoutaro Watanabe, Tim Kretz, and the KIT group for fruitful collaborations.
I am thankful to Hantian Zhang and CERN theory group for the inspiring stay where this document has been prepared.
This work is supported by JSPS KAKENHI Grant Numbers 22K21347, 24K22879, 24K23939 and 25K17385 and Toyoaki scholarship foundation.

\begin{widetext}
\appendix
\section{Comment on the necessity of further $\Xi_b\to \Xi_c$ form factor inputs}
\label{app:FF}
Compared to other transitions the accuracy of the $\Xi_b\to \Xi_c$ form factor determination is not accurate.
Ref.~\cite{Ebert:2006rp} calculated LO IW function based on the relativistic quark model.
In the main text we use $\zeta_s(w)=1+\zeta_s^{(1)}(w-1)+\zeta_s^{(2)}/2(w-1)^2$ where $\zeta_s^{(1)}=-2.27$ and $\zeta_s^{(2)}=7.74$ are set. 
These years the form factor has been calculated in various quark models (see Ref.~\cite{Rui:2025iwa} for a summary table).
However the Lattice result nor experimental data is available currently and the results of QCDSR approach are not fully converging \cite{Lu:2025gol}.
Furthermore the fit result is often available only for $q^2_{\rm{min}}$ and $q^2_{\rm{max}}$ which correspond to $w_{\max}$ and $w=1$ for each.
Unfortunately the $w=1$ data can not constrain $\zeta_s^{(1)}$ and $\zeta_s^{(2)}$.
For instance we fitted the above LO IW function at $\mathcal{O}(1/m_Q,\,\alpha_s)$ to QCDSR constraint \cite{Rui:2025iwa} and obtained a linear relation of parameters as $\zeta_s^{(1)}\simeq-0.2\zeta_s^{(2)}-2.6$\,.
In the kinematic end point, $w_{\rm max}=1.38$ holds for light lepton and we obtain as,
\begin{align}
 \zeta_s(w_{\rm{max}})=1+\zeta_s^{(1)}(w_{\rm{max}}-1)+\zeta_s^{(2)}/2(w_{\rm{max}}-1)^2\simeq1+0.38\zeta_s^{(1)}+0.07\zeta_s^{(2)}\,.
\end{align}
For a better perturbative expansion, the large $\zeta_s^{(1)}$ and $\zeta_s^{(2)}$ are not favorable.
Suppose that $\zeta_s^{(2)}=0$ is set and then we obtain $\zeta_s^{(1)}=-2.6$.
In this case the ratio between the first term and the second term becomes 1.
To make the second term moderate, negative $\zeta_s^{(2)}$ is favored.
We consider three benchmark scenarios $(\zeta_s^{(1)},\,\zeta_s^{(2)})=(-2.6,\,0),\,(-2.15,\,-2)$ and $(-1.8,\,-4),\,$ which are called as $S_1$, $S_2$ and $S_3$, respectively.
It is noted that the sign of $\zeta_s^{(2)}$ is flipped with respect to the result of Ref.~\cite{Ebert:2006rp}.
However it is found that, for $S_2$ and $S_3$ cases, the resulting sum rules are similar to the original ones and $\delta_{\rm NP}\simeq \pm0.01$ are obtained.
In the scenario $S_1$, because of the different $w$ dependence about $20\%$ difference in $a_{\Xi_c}^{V_L T}$ and $a_{\Xi_c}^{V_L T}$ are observed.
Consequently, a larger violation of $\delta_{\rm NP}\simeq\pm0.04$ is observed.
As is mentioned in the main text there is no experimental data nor future prospect in $R_{\Xi_c}$ and thus the impact of this $\pm0.04$ is not clear.
Furthermore it is hard to assign the systematic uncertainty within the QCDSR approach.
We need more independent inputs to fix form factors which enable us to reliably make the prediction and include further higher order corrections \eg $\mathcal{O}(1/m_Q^2)$ terms.

\section{Generic formula of $R_{H_c}/R_{H_c}^{\rm{SM}}$}
\label{app:GF}
The generic R-ratio formulae for mesonic and baryonic decays are shown for the completeness as,
\begin{align}
\label{eq:RD}
 \frac{R_D}{R_D^\textrm{SM}} =
 & ~|1+C_{V_L}|^2  + 1.01|C_{S_R}+C_{S_L}|^2 + 0.84|C_{T}|^2  \nonumber \\[-0.3em]
 & + 1.49\textrm{Re}[(1+C_{V_L})(C_{S_R}^*+C_{S_L}^*)]  + 1.08\textrm{Re}[(1+C_{V_L})C_{T}^*] \,, 
 \\[1em]
\label{eq:RDs}
 \frac{R_{D_s}}{R_{D_s}^\textrm{SM}} =
 & ~|1+C_{V_L}|^2  + 1.05|C_{S_R}+C_{S_L}|^2 + 0.78|C_{T}|^2  \nonumber \\[-0.3em]
 & + 1.53\textrm{Re}[(1+C_{V_L})(C_{S_R}^*+C_{S_L}^*)]  + 1.04\textrm{Re}[(1+C_{V_L})C_{T}^*] \,, 
\end{align} 
\begin{align}
\label{eq:RDst}
 \,\,\,\,\,\frac{R_{D^*}}{R_{D^*}^\textrm{SM}} =
 & ~|1+C_{V_L}|^2  + 0.04|C_{S_R}-C_{S_L}|^2 + 16.0|C_{T}|^2 \nonumber \\[-0.3em]
 &  + 0.12\textrm{Re}[(1+C_{V_L})(C_{S_R}^*-C_{S_L}^*)] -5.17\textrm{Re}[(1+C_{V_L})C_{T}^*]  \,, 
 \\[1em]
 \,\,\,\,\,\frac{R_{D_s^*}}{R_{D_s^*}^\textrm{SM}} =
 & ~|1+C_{V_L}|^2  + 0.04|C_{S_R}-C_{S_L}|^2 + 16.0|C_{T}|^2 \nonumber \\
 &  + 0.11\textrm{Re}[(1+C_{V_L})(C_{S_R}^*-C_{S_L}^*)] -5.39\textrm{Re}[(1+C_{V_L})C_{T}^*]  \,,\\[-0.3em]
\label{eq:RLambda}
 \frac{R_{\Lambda_c}}{R_{\Lambda_c}^{\rm SM}} =  |1+C_{V_L}|^2 
 &+ 0.50 \,\textrm{Re}[ (1 +C_{V_L}) C_{S_R}^*]
 + 0.33 \,\textrm{Re}[ (1 +C_{V_L}) C_{S_L}^* ]
 + 0.52 \, \textrm{Re}[C_{S_L} C_{S_R}^* ]
 \nonumber \\[-0.3em]
 \quad  + 0.32 \, (|C_{S_L}&|^2 + |C_{S_R}|^2) -3.11 \,\textrm{Re}[ (1 +C_{V_L})  C_T^*] + 10.4 \, |C_T|^2\,,\\[1em]
 \frac{R_{\Xi_c}}{R_{\Xi_c}^{\rm SM}} =  |1+C_{V_L}|^2 
 &+ 0.46 \,\textrm{Re}[ (1 +C_{V_L}) C_{S_R}^*]
 + 0.31 \,\textrm{Re}[ (1 +C_{V_L}) C_{S_L}^* ]
 + 0.46 \, \textrm{Re}[C_{S_L} C_{S_R}^* ]
 \nonumber \\[-0.3em]
 \quad  + 0.28 \, (|C_{S_L}&|^2 + |C_{S_R}|^2) - 3.40\,\textrm{Re}[ (1 +C_{V_L})  C_T^*] +  12.3\, |C_T|^2\,.
\end{align}
In our set up we obtain the SM predictions as, 
\begin{align}
    R_{\Lambda_c}^{\rm SM}\simeq0.32,\,\,\,\,R_{D}^{\rm SM}\simeq0.29,\,\,\,\,R_{D^*}^{\rm SM}\simeq0.25,\,\,\,\,R_{\Xi_c}^{\rm SM}\simeq0.25,\,\,\,\,R_{D_s}^{\rm SM}\simeq0.30,\,\,\,\,R_{D_s^*}^{\rm SM}\simeq0.24\,.
\end{align}

\section{Sum rules based on the KIT method}
\label{app:KIT}
In Fig.~\ref{Fig:KIT} we show $\delta_{ij}^{kl}$, $\epsilon_{ij}^{kl}$ and $\delta_{\rm{\rm NP}}^{kl}$ based on the KIT method where a label of the vanishing operator $kl$ is added.
The sum rule coefficient $\alpha$ is $\{0.250,\,0.247,\,0.248,\,0.245\}$ for $\{ \Lambda_c$-$D$-$D^*$, $\Lambda_c$-$D_s$-$D_s^*$, $\Xi_c$-$D$-$D^*$, $\Xi_c$-$D_s$-$D_s^*\}$ combinations.
$\alpha$ is fixed such that $V_LV_L$ and $V_LS_R$ terms are eliminated.
We see that $\alpha$ is about $1/4$ and both methods are almost converging.
For $kl=V_LS_L,\,SS,\,S_RS_L,\,V_LT,\,TT$ the coefficient $\alpha$ is given as $\{0.265,\,0.253,\,0.266,\,0.254\},\,\{0.259,\,0.248,\,0.251,\,0.240\},\,\{0.259,\,0.247,\,0.257,\,0.244\},\,\{0.285,\,0.316,\,0.279,\,0.310\},$
$\,\{0.250,\,0.251,\,0.246,\,0.247\}$, respectively.
We see that the coefficient lists of $kl=V_LS_R$ and $TT$ scenarios are similar to each other.
As a result the $kl=TT$ case is found be as good as the $kl=V_LS_R$ one.
On the other hand in the $kl=V_LT$ scenario the coefficient can be about $0.3$.
In the scenario $\delta_{\rm {\rm NP}}$ is slightly enhanced to be $\simeq \pm0.02$.
We conclude that in the ground to ground combinations the KIT method are as good as the HQ method even with SU(3)$_{\rm{F}}$ rotations.
On the other hand it is fair to comment that the coincidence of $\alpha\simeq0.25$ in both methods is rather accidental.
In the HQ method $\alpha=1/4$ is obtained in heavy quark limit and zero-recoil limit.
However if we divide the available $w$ range into several sub part and consider the last bin around the maximal $w$, the coefficient can be largely deviated from $1/4$ \cite{Endo:2025fke}.

\begin{figure*}[t]
\begin{center}
\includegraphics[width=0.3\textwidth]{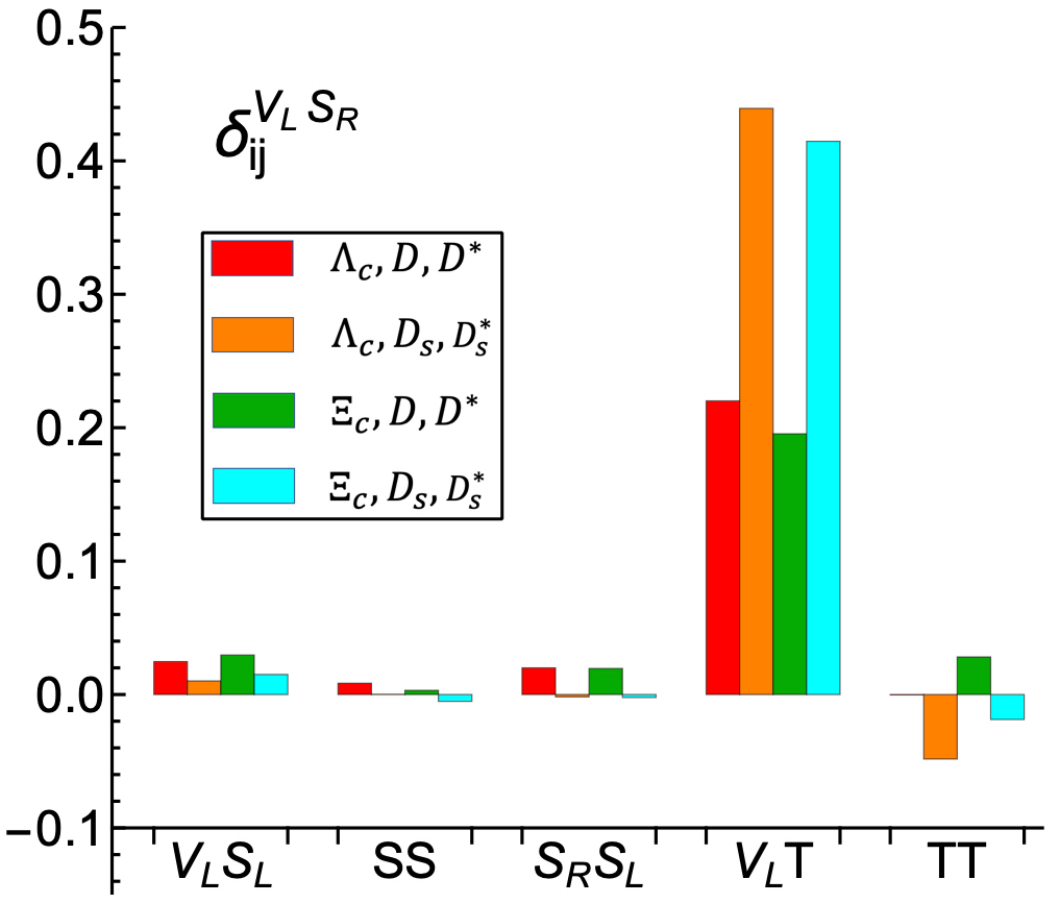}~~~
\includegraphics[width=0.3\textwidth]{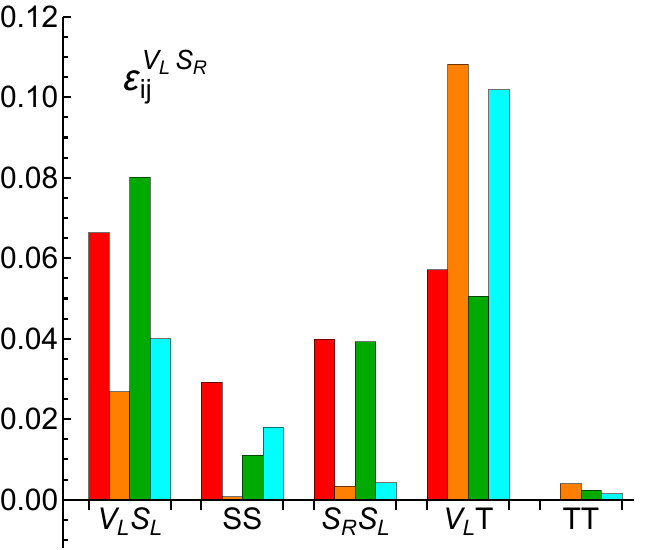}~~~
\includegraphics[width=0.3\textwidth]{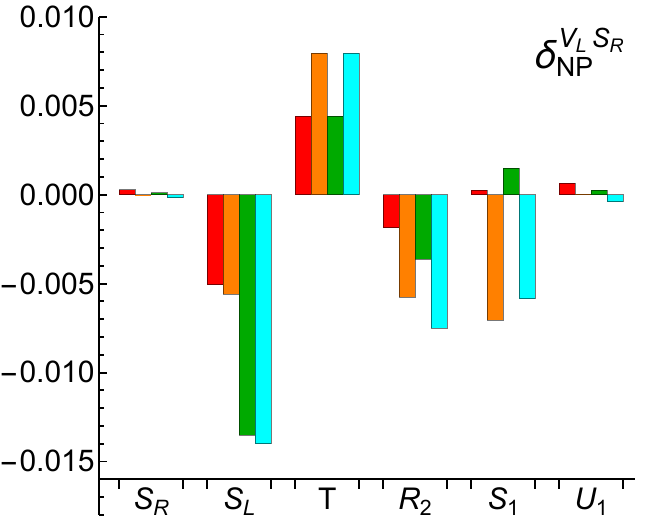}
\caption{
   \label{Fig:KIT}
$\delta_{ij}$ (left), $\epsilon_{ij}$ (middle) and $\delta_{\rm{\rm NP}}$ (right) in the KIT method.
The sum rule coefficient $\alpha$ is determined such that $V_LV_L$ and $V_LS_R$ terms are vanishing from $\delta$.
See the caption of Figs.~\ref{Fig:HQ_delep} and \ref{Fig:HQ_delNP}, for order of operators and scenarios.} 
\end{center}
\vspace{-4mm} 
\end{figure*}

\end{widetext}
\bibliographystyle{utphys}
\bibliography{ref}
\end{document}